\documentclass[reprint,amsmath,amssymb, aps, prb]{revtex4-2}
\usepackage{amsmath}
\usepackage{amsbsy, amssymb}
\usepackage{amsthm}
\usepackage{graphicx}
\usepackage{xcolor}
\usepackage{comment}
\usepackage{ulem}
\usepackage{footmisc}
\usepackage{microtype}
\usepackage{hyperref}
\usepackage{tikz,mathtools}
\usepackage{nccmath}

\renewcommand{\v}{\textbf}
\newcommand{\spin}{\boldsymbol{\sigma}}

\newcommand{\A}{\boldsymbol{\mathcal A}}

\usepackage{mathtools} 

\newcommand{\llangle}{\mathopen{\langle\!\langle}}
\newcommand{\rrangle}{\mathclose{\rangle\!\rangle}}

\newcommand{\sumtikz}[1]{%
\sum\limits_{\mathclap{\vcenter{\hbox{\tikz[baseline=0,scale=1.]{#1}}}}}
}

\newcommand{\bond}{%
  \fill (-.15,0) circle (1.5pt); \fill (.15,0) circle (1.5pt);
  \draw (-.15,0) -- (.15,0);
}

\newcommand{\tri}{%
  \foreach \a in {90,210,330}{\fill (\a:0.173)  circle(1.5pt);}
  \draw (90:0.173)--(210:0.173)--(330:0.173)--cycle;
}

\newcommand{\rhombus}{%
  \coordinate (A) at (0,0);
  \coordinate (B) at (0.3,0);
  \coordinate (C) at (0.45,0.2598);
  \coordinate (D) at (0.15,0.2598076211);
    \draw (A) -- (B) -- (C) -- (D) -- cycle;
  \foreach \P in {A,B,C,D} \fill (\P) circle(1.5pt);
}

\newcommand{\triWide}{%
  \coordinate (A) at (0,0);
  \coordinate (B) at (0.3,0);
  \coordinate (C) at (0.45,0.2598);
    \draw (A) -- (B) -- (C) -- cycle;
  \foreach \P in {A,B,C} \fill (\P) circle(1.5pt);
}

\newcommand{\five}{%
  \coordinate (A) at (0,0);
  \coordinate (B) at (0.3,0);
  \coordinate (C) at (0.6,0);
  \coordinate (D) at (0.45,0.2598076211);
  \coordinate (E) at (0.15,0.2598076211);
  \draw (A) -- (B) -- (C) -- (D) -- (E) -- cycle;
  \foreach \P in {A,B,C,D,E} \fill (\P) circle(1.5pt);
}

\newcommand{\hexagon}{%
  \coordinate (A) at (0,0);
  \coordinate (B) at (0.3,0);
  \coordinate (C) at (0.6,0);
  \coordinate (D) at (0.45,0.2598076211);
  \coordinate (E) at (0.15,0.2598076211);
  \coordinate (F) at (0.45,-0.2598076211);
  \coordinate (G) at (0.15,-0.2598076211);
  \draw (A) -- (G) -- (F) -- (C) -- (D) -- (E) -- cycle;
  \foreach \P in {A,C,D,E,F,G} \fill (\P) circle(1.5pt);
  \foreach \P in {B} \fill (\P) circle(0.8pt);
}

\newcommand{\rhombussix}{%
  \coordinate (A) at (0,0);
  \coordinate (B) at (0.3,0);
  \coordinate (C) at (0.6,0);
  \coordinate (D) at (0.75,0.2598);
  \coordinate (E) at (0.45,0.2598076211);
  \coordinate (F) at (0.15,0.2598076211);
  \draw (A) -- (B) -- (C) -- (D) -- (E) --(F) -- cycle;
  \foreach \P in {A,B,C,D,E,F} \fill (\P) circle(1.5pt);
}

\newcommand{\Triangles}[1][]{%
\mathord{\vcenter{\hbox{\tikz[baseline=-0.2ex,scale=0.7,#1]{\tri}}}}}

\newcommand{\Rhombus}[1][]{%
  \mathord{\vcenter{\hbox{\tikz[baseline=-0.2ex,scale=0.7,#1]{\rhombus}}}}}

\newcommand{\TriWide}[1][]{%
  \mathord{\vcenter{\hbox{\tikz[baseline=-0.2ex,scale=0.7,#1]{\triWide}}}}}

\newcommand{\ksk}[1]{#1}

\begin{document}

\preprint{APS/123-QED}

\title{Exchange Interactions of  a Wigner Crystal in a Magnetic Field and Berry Curvature: \\ Multi-Particle Tunneling through Complex Trajectories}

\author{Kyung-Su Kim}
\altaffiliation{ \href{mailto:kyungsu@illinois.edu}{kyungsu@illinois.edu}
}
\affiliation{Department of Physics and Anthony J. Leggett Institute for Condensed Matter Theory, University of Illinois  Urbana-Champaign, 1110 West Green Street, Urbana, Illinois 61801, USA}

\date{\today}
\begin{abstract}
We study how an out-of-plane magnetic field $B(\v r)$ and a Berry curvature $\Omega(\v k)$ modify the exchange interactions in a two-dimensional Wigner crystal (WC) using a semiclassical large-$r_s$ expansion. 
When only a magnetic field is present, ring-exchange processes arise from multi-particle tunneling through {\it complex} trajectories which constitute {\it complex instanton} solutions of the coordinate-space path integral. 
To leading order in $B$, each exchange constant $J_a$ acquires an Aharonov-Bohm phase along the zero-field tunneling trajectory. 
When a Berry curvature is present, the multi-particle tunneling must be considered in a complexified phase space $(\v r, \v k)$.
To leading order in $\Omega$,  $J_a$ acquires a Berry phase along a {\it purely imaginary} momentum-space trajectory. 
When  $B$ and $\Omega$ are both present, in addition to having both Aharonov-Bohm and Berry phases, the exchange magnitude $|J_a|$ is also exponentially modified due to an effective-mass renormalization.
These effects could be relevant for the WC and proximate phases recently observed in rhombohedral multilayer graphene.
\end{abstract}
\maketitle

\section{Introduction}

The two-dimensional electron gas (2DEG)---Coulomb-interacting electrons confined to move in two dimensions---is governed by the dimensionless parameter $r_s$ characterizing the ratio of interaction to kinetic energy
\begin{align}
    \label{eq:2DEG}
    H_0 &= \sum_{i}  \frac{\v p_i ^2}{2m}
    + \sum_{i<j} \frac{e^2}{4\pi \epsilon} \frac{1}{|\v r_i -\v r_j|},
    \\
\label{eq:r_s}
    r_s &= \frac{e^2/8
\pi\epsilon r_0}{\hbar^2/2mr_0^2}=r_0/a_{\rm B}.
\end{align} 
Here, $r_0 = 1/\sqrt{\pi n}$ is the average interparticle distance (for an electron density $n$)   and $a_\mathrm{B} = { 4\pi \epsilon \hbar^2} /{m e^2}$  is the effective Bohr radius.
The 2DEG forms a Fermi liquid (FL) for $r_s \ll 1$ (weak-coupling regime) and a  Wigner crystal (WC) for $r_s \gg 1$ (strong coupling) \cite{wigner1934interaction, tanatar1989QMC, attaccalite2002QMC, drummond2009QMC, smith2024unified} \cite{footnote:intermediate-phase}.

The magnetism within each such phase of the  2DEG is an interesting and subtle problem that has attracted long-standing attention \cite{tanatar1989QMC, attaccalite2002QMC, drummond2009QMC, hossain2020ferromagnetism, kim2021discovery, falson2022competing}.
In particular, in the WC phase, the magnetism is determined by various multi-spin ring-exchange processes \cite{thouless1965exchange, roger1983RMP, Roger1984WKB, chakravarty1999WC, voelker2001disorder, Katano2000WKB, kim2022interstitial, Ceperley2001exchange} (see \cite{kim2023dynamical} for the most recent study).
The effective spin Hamiltonian is written as a sum over various ring-exchange interactions:
\begin{align}
\label{eq:WC exchanges}
    &H_{\rm{eff}} = -\sum_{a} (-1)^{P_a}\, {J_a} \,  \mathcal{\hat P}_a 
    \nonumber
     \\
    &= 
    \sumtikz{\bond}(J_2 \, \mathcal{\hat P}_{2}+ J_2^{*}\,\mathcal{\hat P}_2 ^{-1})
    -  \sumtikz{\tri}(J_3 \, \mathcal{\hat P}_3+ J_3^{*} \, \mathcal{\hat P}_3 ^{-1})
    \\ &
    +  \sumtikz{\rhombus}
    (J_4 \, \mathcal{\hat P}_4 + J_4^{*}  \, \mathcal{\hat P}_4 ^{-1})
    - \sumtikz{\five} ( J_5 \, \mathcal{\hat P} _ 5+ J_5^{*} \, \mathcal{\hat P}_5^{-1})
    \nonumber
    \\& 
    + \sumtikz{\hexagon}( J_6\, \mathcal{\hat P}_6 + J_6^{*} \, \mathcal{\hat P}_6 ^{-1})
    + \sumtikz{\rhombussix}( J_6' \, \mathcal{\hat P}_6 + J_6'^{*} \, \mathcal{\hat P}_6 ^{-1})
    + \cdots, \nonumber 
\end{align}
where $P_a$ refers to an $n_a$-particle exchange that permutes spins in each polygon in \eqref{eq:WC exchanges} in a counter-clockwise direction  and $(-1)^{P_a} = (-1)^{n_a+1}$ is the sign of the permutation.
$\mathcal{\hat P}_a $ is the permutation operator for $P_a$, which could in turn be expressed in terms of spins $\v S$ [as in (\ref{eq:two-particle exchange}-\ref{eq:h_ijkl})].
$J_a$ is the exchange constant for $P_a$  that is in general complex when the magnetic field $B$ or Berry curvature $\Omega$ is present ($J_a^*$ is the complex conjugate).
When $B=\Omega =0$,  the  semiclassical large-$r_s$ approximation gives \cite{chakravarty1999WC, Katano2000WKB, voelker2001disorder, kim2023dynamical}
\begin{equation}
    \label{eq:exchange-constant-introduction}
    J_a[B =0,\!\Omega\!=\!0]  
    =  J_a^{(0)} (E_h, r_s) \equiv  
    E_h  
     \frac {A_a^{(0)} \! \sqrt{\!\tilde S_a^{(0)}}\!} { r_s^{5/4} \! \sqrt {2\pi } }
     e^{-\sqrt{r_s}\tilde  S_a^{(0)}}\!,
\end{equation}
where  $E_h \equiv  \frac{e^2}{4\pi\epsilon a_{\rm B}} $ is the effective Hartree energy.
$\tilde S_a^{(0)}$ and $A_a^{(0)}$ are, respectively, the normalized action and  fluctuation determinant shown in Table \ref{table:results}.

How does an  out-of-plane magnetic field $B(\v r)$ and a Berry curvature $\Omega(\v k)$ affect these exchange interactions?
We address this  in a 2DEG   with a  (possibly non-uniform) $B(\v r)$ 
\footnote{
    $B(\v r) $ can also be regarded as an emergent magnetic field arising from topologically non-trivial spin or pseudo-spin texture in a real space, as discussed in \cite{morales2024magic}. 
    The spatial modulation of $B(\v r)$ at microscopic WC lattice scale is natural in this setting.
}
and/or $\Omega(\v k)$ inherited from the  parent Bloch band.
Throughout this paper, we assume a small effective $g$-factor ($g\ll 1$) and ignore Zeeman splitting \ksk{\cite{footnote:g-factor}}. 
Working deep in a WC regime (large $r_s$), we derive the asymptotic expression for $J_a$ for three settings: 
(i) $B\neq 0$ and $\Omega =0$;
(ii) $B=0$ and $\Omega \neq 0$;
and (iii) $B\neq 0$ and $\Omega  \neq 0$.
In cases (i-ii),  the exchange interactions acquire geometric phase factors which in turn generate chiral spin interactions $\v S_i \cdot (\v S_j \times \v S_k)$ and modulate the Heisenberg couplings.
In case (iii), in addition to having phase factors, the exchange magnitude $|J_a|$ is also exponentially renormalized, providing a knob for an orders-of-magnitude tuning of exchange scale with $B$.
Before turning to the technical details, we  summarize our main findings below.

\setlength{\tabcolsep}{10pt} 
\begin{table}[t]
\centering
    \begin{tabular}{lrrrr}
    \hline
\!\!\!\!\!\!\!\!Process & $ \tilde S_a^{(0)}$ & $A_a^{(0)}$ & $\tilde \Sigma^{(r)}_a$ & $\tilde \Sigma^{(k)}_a$ \\
\hline\hline
$J_2$  & 1.63 &   1.28  &  1.43  &  -0.245  \\
$J_3 $ & 1.53 & 1.18 & 3.03 & -0.106 \\
$J_4 $ & 1.65 & 1.32 & 4.62 & -0.082 \\
$J_5 $ & 1.90 & 1.72 & 6.22 & -0.077 \\
$J_6 $ & 1.79 & 1.62 & 10.87 & -0.042 \\
$J_6'$ & 2.11 & 2.48 & 7.79 & -0.069 \\
\hline\hline
\end{tabular}
    \caption{$\tilde S_a^{(0)}$ and $A_a^{(0)}$: Dimensionless action and fluctuation determinant that determine exchange constants for $B=\Omega =0$ \eqref{eq:exchange-constant-introduction}; 
    $\tilde \Sigma^{(r)} _a $: dimensionless area enclosed by the real-space trajectory ${\bf \tilde {\v r}}^{(0)}_a(\tilde \tau) $ \eqref{eq:area}; 
    $ \tilde \Sigma^{(k)} _a$: dimensionless area enclosed by momentum-space trajectory 
    ${\bf \tilde {\v k}}^{(0)}_a(\tilde \tau )$ \eqref{eq:area momentum}. 
    These values are calculated from the semiclassical trajectories---solutions to \eqref{eq:real instanton equation}---by discretizing  time into $N_{\rm time}=50$ slices and allowing  $N_{\rm move}=30,40,...,90$ electrons to move during the tunneling process, and then extrapolating  to $N_{\rm move} \! \to \infty $.
    The details of the numerical calculation can be found in   \cite{kim2023dynamical}.  
    }
    \label{table:results}
\end{table}

\subsection{Summary of Main Results}

\begin{figure}[t]
    \centering
    \includegraphics[width=\linewidth]{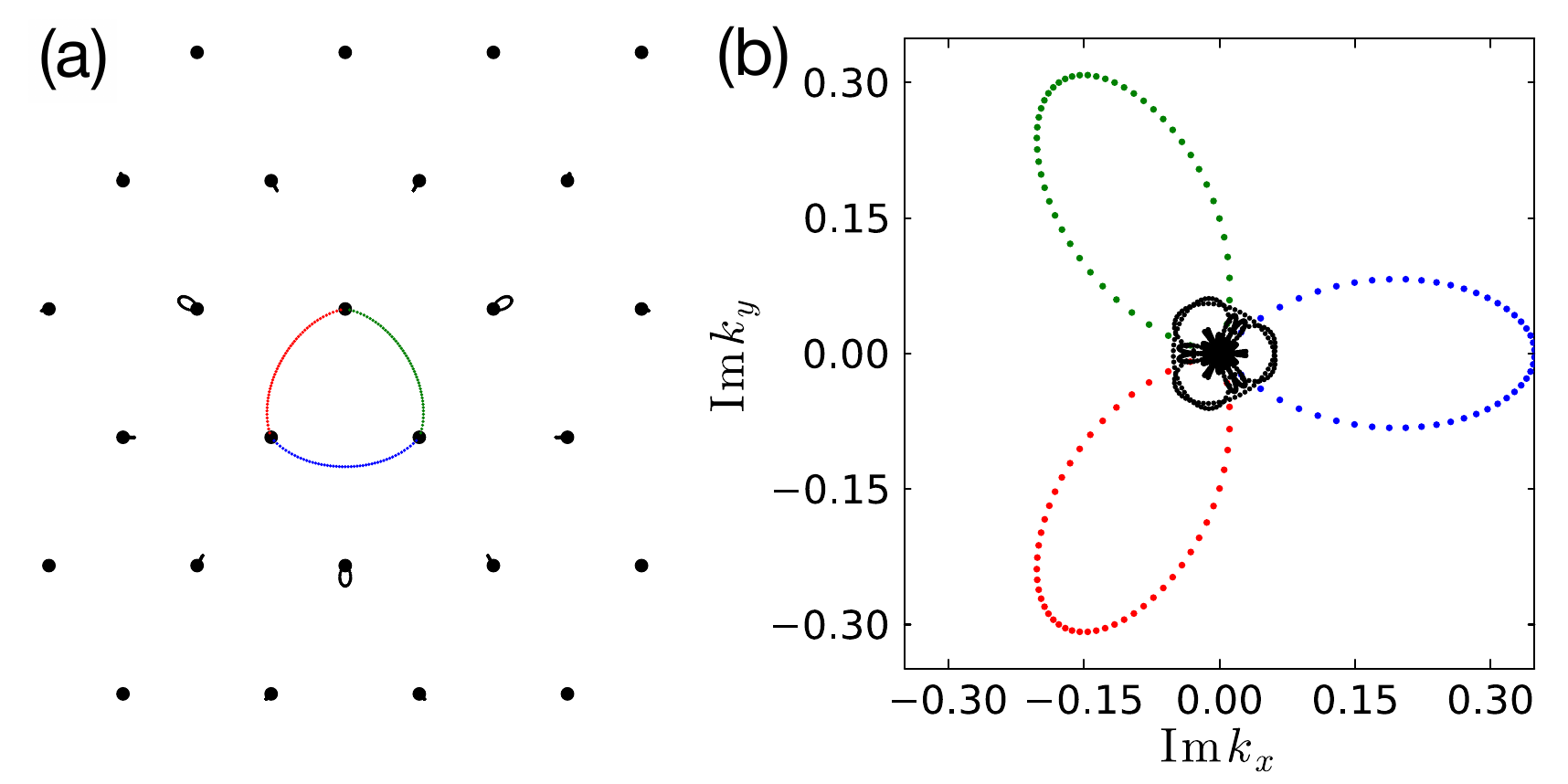}
    \caption{\ksk{Leading-order tunneling trajectory for the $J_3$ process in phase space $(\tilde{\mathbf r}_i^{(0)},\tilde{\mathbf k}_i^{(0)})$, obtained from Eq. \eqref{eq:Hamilton's eq at 0th order}. (a) Real-space trajectory $\tilde{\mathbf r}_i^{(0)}(\tau)$. (b) Purely imaginary momentum-space trajectory $\tilde{\mathbf k}_i^{(0)}(\tau)= i\,\dot{\tilde{\mathbf r}}_i^{(0)}(\tau)$. Colors indicate the trajectories of the three permuting particles.  $\tilde{\Sigma}_a^{(r)}$ and $\tilde{\Sigma}_a^{(k)}$ are the dimensionless areas enclosed by the tunneling trajectory in real space (a) and in momentum space (b), respectively.}
    }
     \label{fig:tunneling}
\end{figure}

(i) We first consider the case with only a  magnetic field, $B(\v r ) = \nabla_{\v r } \times \v A (\v r) =  B_0  + \delta B(\v r )$, where 
$B_0$ is the average field
\cite{footnote:modulation}.
We solve the multi-particle tunneling problem in the $r_s\to \infty$ limit at a fixed ``Landau-level filling factor''
\begin{align}
\label{eq:Landau filling factor}
    \nu \equiv n\cdot 2\pi \ell_{ B_0}^2 = 2 (\ell_{ B_0}/r_0)^2 = O(1),
\end{align}
where $\ell_{B_0}\equiv \sqrt{\hbar /e  B_0}$ is the magnetic length  \cite{footnote:LL-filling-factor}.
(Throughout this paper, we assume  $ B_0 \geq 0$ without loss of generality.)
We assume $\nu$ to be of order 1 and use $(\nu \sqrt{r_s})^{-1}\ll 1$ as a small expansion parameter---\ksk{which we call the ``small cyclotron condition'' }
\cite{footnote:small_parameter}.
To leading order, exchange constants are modified from their zero-field values by an Aharonov-Bohm phase factor:
\begin{fleqn}
    \begin{align}
        \label{eq:exchange1}
        &J_a ^{(0)} \rightarrow J_a [B (\v r)]= J_a ^{(0)} e^{i \phi_a [B (\v r)]},
        \\
        \label{eq:Aharonov-Bohm}
         \phi_a [B (\v r)]  \!=\!  & - \frac{e}{\hbar} \!\sum_i \! \!\int \! d\v r^{(0)}_{i,a} \!\cdot\! \v A (\v r^{(0)}_{i,a}) \!=\!   
        - \frac{e }{\hbar} \! \iint_{S_a^{(r)}} \!\!\!\!\! B (\v r) d^2\v r  ,\!\!
    \end{align} 
\end{fleqn}
where $\v r_{i,a}^{(0)}(\tau)$ is the (position-space) tunneling trajectory of the $i$th electron in the absence of a magnetic field [see Fig. \ref{fig:tunneling} (a) for the illustration of the tunneling trajectory for the $J_3$ process] and $S_a^{(r)}$ is an oriented surface enclosed by the union of trajectories $\v r_{i,a}^{(0)}(\tau)$ for all $i$.
When the magnetic field is  uniform, $\delta B(\v r ) = 0 $,  $\phi_a[B(\v r)]$ further simplifies to 
\begin{align}
    \label{eq:Aharonov-Bohm_uniform}
    & \phi_a [B_0]= - \frac{e B}{\hbar} \Sigma_a^{(r)} = - \frac{2}{\nu}  \tilde \Sigma_a^{(r)} ,
    \\
    \label{eq:area}
    & \Sigma_a^{(r)} \equiv \iint_{S_a^{(r)}} \!\! d^2 \v r  = \sum_i \frac  1 2  \int \v r _{i,a}^{(0)} \times d \v r _{i,a}^{(0)} \equiv r_0^2 \cdot \tilde \Sigma^{(r)} _a,
\end{align} 
where $\Sigma_a^{(r)}  $ is the (signed) area of $S_a^{(r)}$, which is positive for a counter-clockwise motion. 
 $\tilde \Sigma_a^{(r)}$ is the dimensionless area (shown in Table \ref{table:results}) which is independent of  $r_s$ and $\nu$. 

\ksk{  
We note that the effects of a magnetic field on exchange constants were previously studied in Ref. \cite{okamoto1998magnetism, hirashima2001multiple}. 
In particular, our calculations of the dimensionless area $\tilde \Sigma_a^{(r)}$ in Table \ref{table:results} agree with the analogous quantity $s_n^{(0)}/\triangle$ of Ref. \cite{hirashima2001multiple} up to an appropriate numerical factor.
However, the theoretical analysis in \cite{hirashima2001multiple} did not properly account for the complexified path integral needed to obtain the correct asymptotic form of the exchange constants. 
As we will later show, the relevant tunneling trajectory is obtained by solving the complex instanton equations \eqref{eq:equation of motion}.
As a result, although the leading-order expression (namely, the phase factor) in Ref. \cite{hirashima2001multiple} coincides with our result, their subleading corrections derived there are not valid.
}

\smallskip

(ii) When only a  Berry curvature is present, $\Omega(\v k) = 
\nabla_{\v k} \times \boldsymbol  {\mathcal{ A }} 
(\v k ) =  \Omega_0 + \delta \Omega (\v k) $, 
where $ \Omega _0$ is the average Berry curvature
\cite{footnote:modulation},
we solve the problem at large $r_s$ for a fixed 
\begin{align}
\label{eq:alpha}
    \alpha \equiv  r_s  \nu_{\Omega} =  \frac{2 \Omega_0}{a_B^2 r_s} =O(1).
\end{align}
We assume $\alpha = O(1)$ and define the Chern filling factor,
\begin{align}
\label{eq:Chern filling factor}
    \nu_{ \Omega} \equiv  \Omega_0 \cdot 2\pi n = {\rm sgn}( \Omega_0)\cdot  2 (\ell_{  \Omega_0}/r_0)^2 ,    
\end{align}
with the Berry length  $\ell_{\Omega_0} \equiv \sqrt{|\Omega_0|}$.
To leading order in $\alpha /\!\sqrt {r_s}= \nu_{\Omega } \sqrt r_s$, $J_a$ acquires a Berry phase 
along the $\Omega =0$ momentum-space tunneling trajectory $\v k_{i,a}^{(0)}(\tau)$:
\begin{align}
\label{eq:exchange2}
    &J_a ^{(0)} \rightarrow J_a [\Omega(\v k )]= J_a ^{(0)} e^{i \gamma_a [\Omega(\v k )]},
\end{align}
\begin{align}
     \label{eq:Berry phase}
     \gamma_a[\Omega(\v k )] &= \sum_i \! \oint d\v k^{(0)}_i \cdot \boldsymbol{\mathcal A}  (\v k^{(0)}_i) =   \!
    \iint_{S_a^{(k)}} \!\! \Omega (\v k) d^2\v k   .
\end{align} 
The momentum space trajectory for the $J_3$ process is illustrated in Fig. \ref{fig:tunneling} (b).
Here, $S_a^{(k)}$ is an oriented surface enclosed by the union of  momentum-space trajectories $\v k_{i,a}^{(0)}(\tau)$ for all $i$.
When the Berry curvature is  uniform, $\delta \Omega(\v k) = 0$,
\begin{align}
     \label{eq:Berry-phase_uniform}
    & \gamma_a[\Omega_0]=   \Omega_0\cdot  \Sigma_a^{(k)} = \frac \alpha 2   \tilde \Sigma_a^{(k)} ,
    \\
    & \Sigma_a^{(k)} \equiv   \sum_i \frac 1 2 \oint \v k_{i,a}^{(0)} \times  d\v k_{i,a}^{(0)} 
    \nonumber \\
    &= -\sum_i \frac 1 2 
    \oint \mathcal{I}m  \v k_{i,a}^{(0)} \times d \mathcal{I}m \v k_{i,a}^{(0)}
    \equiv \frac 1 {a_B r_0} \tilde \Sigma _a^{(k)},
    \label{eq:area momentum}
\end{align} 
where $\Sigma_a^{(k)}$ is the signed area of $S_a^{(k)}$ and $\tilde \Sigma_a^{(k)}$ is  the dimensionless momentum-space area independent of $r_s$ and $\nu_{\Omega}$ (Table \ref{table:results}).
Interestingly, at leading order, the momentum  $\v k^{(0)}_a(\tau)$ is purely imaginary, and thus $\Sigma_a^{(k)} < 0$ for a counter-clockwise motion in coordinate space.

\smallskip

(iii) Finally, we consider the case with both $\Omega (\v k)$ and a {\it uniform} magnetic field $B(\v r) = B_0$ 
\footnote{When the magnetic field is spatially non-uniform, the effective mass has a spatial dependence $m^*(\v r)$ \eqref{eq:effective band mass}, which modifies semiclassical equations of motion. 
More specifically, the right-hand-side of (\ref{eq:Hamilton equation for k}) must be added by 
\begin{align*}
    -i  \frac{\tilde k^2_i }{2 [\tilde m^*({\bf \tilde {r}}_i)]^2} \nabla_{{\bf \tilde {r}} _i } \tilde m^*({\bf \tilde {r}}_i ),
\end{align*}
where  ${\bf \tilde {r}}_i$ and  ${\bf \tilde k}_i $ are dimensionless coordinates and momenta defined later and  $\tilde m^*({\bf \tilde {r}}) \equiv m^*(r_0  {\bf \tilde {r}})/m_0$ with $m_0$ being the spatially average mass.
}.
In this case, the Aharonov-Bohm and Berry phases contribute additively
\begin{fleqn}
    \begin{align}
    \label{eq:AB and Berry}
        J_a ^{(0)} \rightarrow J_a [B_0,\Omega(\v k) ] =  J_a ^{(0)} (E_h^*, r_s^*)   e^{i  \phi_a [B_0] + i \gamma_a [ \Omega(\v k)] }\!
    \end{align}
\end{fleqn}
but crucially with a renormalized $\alpha \to \alpha ^* \equiv r_s^* \nu_{\Omega}$  \eqref{eq:alpha} and  exchange magnitude $J_a ^{(0)} (E_h, r_s)  \to J_a ^{(0)} (E_h^*, r_s^*)$ 
[$r_s^*$ and $E_h^*$ are to be defined].
This is because the coupling of the orbital magnetic moment $M_z$ to the magnetic field changes the band dispersion relation by 
$\Delta \varepsilon =  -  B _0  M_z(\v q= \v k + \v Q_0 )$  \cite{xiao2010berry}, where $\v k$ is the small deviation from the band minimum $\v Q_0$  and  $\v q = \v k + \v Q_0$ is the  crystal momentum.
\ksk{In this work, we restrict ourselves to the case where only a single valley (or band) near $\v Q_0$ is active at low energy. (See the later discussion in Sec. III for more detail.)}
Within the effective mass approximation ($\v k \sim 0$), this amounts to the  effective-mass renormalization 
\footnote{
    In this work, the effective  mass tensor is assumed to be isotropic.
}, 
\begin{align}
\label{eq:effective band mass}
    m \to m^* = \left (\frac 1 m -\frac {B _0} {\hbar^2}  \frac {d^2 M_z}{dq^2} \bigg \vert_{\v q=\v Q_0} \right )^{-1}.  
\end{align}
This in turn leads to the renormalization of $a_B$, $r_s$, $E_h$ and $\alpha$: 
\begin{align}
\label{eq:renormalized parameters}
    a_B ^* = \frac {4\pi \epsilon \hbar^2} {m^* e^2},\  r_s^* = \frac {r_0} {a_B^*}, \  E_h^* = \frac {e^2}{4\pi \epsilon a_B^*},\   \alpha ^* =r_s^* \nu_\Omega.
\end{align}
Since $|J_a| \propto \exp[-\sqrt {r_s^*} \tilde S_a^{(0)}]$  \eqref{eq:exchange-constant-introduction}, this means that 
even a modest field-induced  renormalization of $m^*$ can exponentially enhance or suppress the magnetic energy scale.

\section{Effects of a  magnetic field}

We first derive the exchange interactions of the WC in  a  magnetic field, $ B (\v r ) = \nabla_{\v r} \times \v A (\v r)  = B_0  +\delta B( {\v r})$,
\begin{align}
    \label{eq:2DEG with B}
    H = \sum_{i}  \frac{\left [\v p_i + e \v A(\v r_i) \right ]^2}{2m}
    + \sum_{i<j} \frac{e^2}{4\pi \epsilon} \frac{1}{|\v r_i -\v r_j|},
\end{align}
generalizing the analogous derivation for $B=0$ (see e.g., \cite{kim2023dynamical, voelker2001disorder}). 
Consider a single tunneling process $P_a$ that cyclically permutes $n_a$ electrons at WC sites $\v R_{i_k}$ with spins $\sigma_k$ to $\v R_{i_{k+1}}$ ($i_{n_a + 1} \equiv  i_1$)
\begin{align}
    (\v R_{i_k}, \sigma_{k}) 
    \underset{P_a}{\rightarrow} 
    (\v R_{i_{k+1}}, \sigma_{k}), \ \ k=1,\cdots, n_a. 
\end{align}
The corresponding matrix element for the  effective Hamiltonian   is 
\begin{equation}
\label{eq:matrix element}
    \langle  P_a  \spin   | h_a  |  \spin \rangle  = - J_a[B], 
\end{equation}
where $\spin  \equiv ( \sigma_1, \sigma_2, \cdots, \sigma_N)$ is a spin configuration with $\sigma_i$ at a WC site $ \v R_i $ and $P_a \spin  \equiv ( \sigma_{P^{-1}_a(1)}, \cdots, \sigma_{P^{-1}_a(N)})$ is the permuted configuration. 
Writing $h_a$ in second-quantized form in terms of  $f_{i, \sigma}^{\dagger}$ and $f_{i, \sigma}$, the electron creation and annihilation operators at  $\v R_i$,
\begin{fleqn}
    \begin{align}
    \label{eq:single exchange}
        &h_a \! = \! - J_a \! \sum_{\{\sigma_i \} }\!\!
        \left (
        f^{\dagger}_{i_2, \sigma_1}  f^{\dagger}_{i_3,\sigma_2} \!\!\cdots \! f^{\dagger}_{i_1, \sigma_{n_a}} \! \right )\!\!  
        \Big ( f_{i_{n_a},\sigma_{{n_a}}} \!\!\cdots\! f_{i_2,\sigma_{2}} f_{i_1,\sigma_{1}} \!\Big )
        \nonumber \\
        & \ \ \, = - (-1)^{P_a}  J_a \, \mathcal{\hat P}_a  ,
    \\
    \label{eq:permutation operator}
        &\mathcal{\hat P}_a  \equiv   \sum_{\{\sigma_i \} }   f^{\dagger}_{i_{n_a},  \sigma_{n_a -1 }} f_{i_{n_a},\sigma_{n_a}} \cdots 
        f^{\dagger}_{i_1, \sigma_{n_a}} f_{i_1,\sigma_1} \nonumber 
        \\
        & \ \ \ \, =\mathcal{\hat P}_{i_1 i_2 } \mathcal{\hat P}_{i_2 i_3 } \cdots \mathcal{\hat P}_{i_{n_a-1 } i_{n_a} },  
    \end{align}
\end{fleqn}
where  $(-1)^{P_a}= (-1)^{n_a+1 }$ is the parity of $P_a$ \footnote{
    The operators in the first line of \eqref{eq:single exchange} are `normal-ordered' such that electrons with spins $\sigma_1, \sigma_2, \cdots, \sigma_{n_a}$ are annihilated in order and then created in a reverse order. 
    See relevant discussions, e.g., in \cite{kim2024itinerant, sakurai2020modern}.
}. 
The permutation operator $\mathcal{\hat P}_a$ can in turn be written as
\begin{align}
    \label{eq:two-particle exchange}
     \mathcal{\hat P}_{ij} &= 2\, \v S_i \!\cdot\! \v S_j + \frac 1 2,
    \\
    \label{eq:three-particle exchange}
    \mathcal{\hat P}_{ijk } & 
    =
    \v S_i \!\cdot\! \v S_j + \v S_j\! \cdot \!\v S_k+ \v S_k \!\cdot\! \v S_i - 2i \chi_{ijk } + \frac 1 4,
    \\
    \label{eq:spin chirality}
    \chi_{ijk} &\equiv \v S_i \cdot \left (  \v S_j  \times \v S_k \right ),
    \\ 
    \label{eq:four-particle exchange}
 \mathcal{\hat P}_{ijkl }
    &= 
   \frac 1 2 \left ( \v S_i \cdot \v S_j + \cdots \right )
   -i \left ( \chi_{ijk } + \chi_{jkl }+\chi_{kli }+\chi_{lij } \right )
   \nonumber \\ 
   & \   + 2\, h_{ijkl} + \frac 1 8,
   \\
   \label{eq:h_ijkl}
    h_{ijkl} &\equiv  
    (\v S_i \cdot \v S_j  )(\v S_k \cdot \v S_l)
   +(j \leftrightarrow l) - (j \leftrightarrow k),
\end{align}
etc. Here,  the first term in  $\mathcal{\hat P}_{ijkl }$ is the sum over six distinct Heisenberg couplings between any pair of spins among $\{i,j,k,l\}$ and 
$\chi_{ijk} $ is the spin-chirality operator.
Summing over all $h_a$ \eqref{eq:single exchange} reproduces the effective  Hamiltonian \eqref{eq:WC exchanges}.

\subsection{Semiclassical Limit $r_s \to \infty$}

We now obtain the  large-$r_s$ asymptotic form of $J_a[B]$.
This is done by first obtaining the asymptotic behavior of multi-particle propagators at an intermediate time scale 
\begin{equation}
    \label{eq:intermediate time scale}
    1/ E_{\rm Debye} \ll  \Delta \tau \ll 1/|J_a|,
\end{equation}
where 
$E_{\rm Debye} \equiv e^2/(4 \pi \epsilon a_{\rm B} r_s^{3/2})$ sets the zero-point energy scale (per particle) of the WC. [The meaning of such time scale will become clearer later. See e.g., the discussion right above \eqref{eq:exchange constant intermediate}.]
At large $r_s$, the low-energy Hilbert space is spanned by the $2^N$ spin configurations $\spin \equiv (\sigma_i)$ on the triangular WC sites $\v R = (\v R_1, \cdots , \v R_N )$:
\begin{align}
    | \v R , \spin \rangle  \equiv  |(\v R_i,\sigma_i) \rangle.     
\end{align}
Projecting onto this subspace and expanding the off-diagonal element of the exponential to first order 
\footnote{
    The tilde means that the left-hand side is asymptotically equivalent to the right-hand side for $ 1/E_{\rm Debye} \ll  \Delta \tau \ll 1/|J_a|$
},
\begin{align}
    \label{eq:diag} 
    \langle \v R , \spin   | e^{- \Delta \tau  H} |\v R ,   \spin \rangle 
    &\sim 
    |\psi(\v R )|^2  e^{-\Delta \tau E_0},
    \\*
    \label{eq:offdiag}
    \langle P_a ( \v R ,  \spin )  | e^{- \Delta \tau   H }| \v R  , \spin   \rangle
    &\sim |\psi(\v R )|^2 e^{-\Delta \tau E_0} 
     \Delta \tau  J_a,
\end{align}
where $E_0 $ is the average energy of  classically degenerate states $|\v R \!,\spin  \rangle$ and 
$|\psi(\v R )|^2  = |\psi(P_a \v R )|^2 $ is the probability density of electrons occupying the WC sites $\v R $.
Taking the ratio between the two propagators, we obtain
\begin{equation}
    \label{eq:exchange constant 1}
    J_a \sim \Delta \tau ^{-1 } \frac{\langle P_a (\v R \!, \spin) | e^{-\Delta \tau H} | \v R \!,\spin \rangle}{\langle \v R \!,\spin  | e^{-\Delta \tau H} | \v R \!,\spin \rangle}  .
\end{equation}

The propagators can be represented as a path integral
\begin{fleqn}[0pt]
    \begin{align}
    \label{eq:path integral}
    \!\!\!& \langle {\v R^{ \prime }}  ,\spin'  | e^{-\Delta \tau H} |\v R , \spin  \rangle
    = 
    \delta_{\spin, \spin' }  
        \int_{  { \v r}(0) = {\v R }}^{ {\v r}(  \Delta  \tau) = 
      {\v R^{ \prime }}} \!\! D  {\v r}( \tau) e^{-  S} 
    \nonumber 
    \\
    & = \delta_{\spin, \spin'}
        \int_{ {\bf  \tilde { \v r}}(0) = {\bf \tilde {\v R}} }
        ^{ {\bf \tilde {\v r}}(  \Delta \tilde \tau) = 
    {{\bf \tilde {\v R}}^{\prime  }}}  D {\bf \tilde {\v r}}(\tilde \tau) e^{- \sqrt{r_s} \tilde S},
    \\
    \label{eq:action-unnormalized}
    & S[{\v r}(\tau)] 
    = \int_0^{\Delta \tau } d\tau \Bigg [ \sum_i \left \{ \frac { m  } {2\hbar^2}
    \left ( \frac { d{\v r}_i}{d\tau }\right )^2 
     +\frac{ ie}{\hbar} 
       \frac { d{\v r}_i}{d\tau } \cdot  {\v A} ( {\v r}_i) \right \} 
    \nonumber \\
    &\ \ \ \ \ \ \ \ \ \ \  + \frac{e^2 }{4\pi \epsilon }  \sum_{i<j} \frac 1 {|\v r_i - \v r_j|} -  V_0  \Bigg ] 
    = \sqrt r_s  \tilde S[{\bf {\tilde {\v r}}}(\tilde \tau)]
    ,
    \\
    &
    \label{eq:action}
    \tilde S[{\bf {\tilde {\v r}}}(\tilde \tau)] 
    \! = \! \int_0^{\Delta \tilde \tau}  \!\!\!\!\!d\tilde\tau \! \left [ \sum_i \left \{ \frac {\dot{\bf {\tilde {\v r}}}_i^2 } 2    + \! \frac{2 i}{\nu\sqrt{r_s} }  \dot {\bf {\tilde {\v r}}}_i \! \cdot \!  {\bf \tilde {\v A}} ({\bf {\tilde {\v r}}}_i) \! \right \} \! + \! V ({\bf {\tilde {\v r}}}) - \tilde V_0 \right ]\! , \!
    \\
    \label{eq:dimensionless vector potential}
    &  
    {\bf \tilde {\v A }}({\bf {\tilde {\v r}}}) 
    \!=\!
    \frac 1 {r_0 B_0} \v A(\v r),
    \ \tilde B({\bf {\tilde {\v r}}}) 
    \!=\! \nabla_{{\bf {\tilde {\v r}}}} \!\times\! {\bf \tilde {\v A}}({\bf {\tilde {\v r}}})\!=\! 
    1 \!+\! \frac {\delta B(r_0 {\bf {\tilde {\v r}}})}  {B_0} 
 ,
    \\
    \label{eq:Coulomb}
    &
    V({\bf {\tilde {\v r}}}) \equiv \sum_{i<j} \frac{1}{|{\bf {\tilde {\v r}}}_i - {\bf {\tilde {\v r}}}_j|}.
    \end{align}
\end{fleqn}
Here, ${\bf { {\v r}}}( \tau) \equiv ({\bf { {\v r}}}_1( \tau),\cdots , {\bf {{\v r}}}_N( \tau))$ is the collective coordinates of all electrons.
In the second line of \eqref{eq:path integral}, the Euclidean action $ S$ is rescaled to make the $r_s$ dependence manifest by introducing the dimensionless coordinates and imaginary time, ${\bf {\tilde {\v r}}} \equiv {\v r}/r_0$ and $\tilde \tau = \tau E_{\rm Debye}$, where 
$E_{\rm Debye} \equiv e^2/(4 \pi \epsilon a_{\rm B} r_s^{3/2})$. 
${\bf \tilde {\v R}}  = ({\bf \tilde {\v R}}_1 , \cdots ,{\bf \tilde {\v R }}_N )$ is the WC configuration in the normalized coordinate.
The $\delta_{\spin, \spin '}$ factor appears because each electron's spin is conserved under the  dynamics of the Hamiltonian \eqref{eq:2DEG with B}.
The minimum (classical) Coulomb energy corresponding to the WC configuration $V_0  \equiv  \frac {e^2}{4\pi \epsilon }  {\rm min}_{{\v r}}V({\v r})$ is subtracted in \eqref{eq:action-unnormalized} for convenience; correspondingly, $\tilde V_0 =  (\frac {e^2}{4\pi \epsilon r_0})^{-1} V_0  = {\rm min}_{{\bf \tilde {\v r}}}V({\bf \tilde {\v r}})$ is the minimum dimensionless Coulomb energy.  
In \eqref{eq:action}, the dot indicates a derivative with respect to the dimensionless time: $\dot{\bf \tilde {\v r }}_i \equiv d{\bf \tilde {r }}_i /d\tilde{\tau}$.
$\nu = nh /eB_0$  \eqref{eq:Landau filling factor}.
${\bf \tilde {\v A}}(\tilde {\v r}) $  and $\tilde B (\tilde {\v r})$ are the dimensionless vector potential and magnetic field, respectively.
The Coulomb interaction  $ V$ is calculated using the standard Ewald method. 
\footnote{
    As usual, it is assumed that a uniform neutralizing positively-charged background is  present, which cancels the diverging contribution in  \eqref{eq:Coulomb}.
}

At large $r_s$, the dominant contributions to each propagator in  \eqref{eq:exchange constant 1} come from small oscillations around the classical WC configurations together with multi-particle tunneling (or instanton) events connecting one WC configuration to another.  
Since $1/E_{\rm Debye}$ is the typical size of instantons in imaginary time and $1/|J_a|$ is the average distance between them, at most one instanton  can exist within an intermediate time scale $\Delta \tau$ \eqref{eq:intermediate time scale}.
Therefore, 
\begin{align}
    \label{eq:exchange constant intermediate}
    J_a &\sim \Delta \tau ^{-1 } \frac{\langle  P_a (\v R , \spin)  | e^{-\Delta \tau H} |  \v R , \spin \rangle|_{\text{1-inst}} }
    {\langle  \v R , \spin | e^{-\Delta \tau H} |  \v R , \spin \rangle|_{\text{0-inst}} } .
\end{align}
The denominator of \eqref{eq:exchange constant intermediate} can be explicitly evaluated at large $r_s$ by considering harmonic fluctuations $\delta {\bf {\tilde {\v r}}}(\tilde \tau)$  around a WC configuration: ${\bf {\tilde {\v r}}}(\tilde \tau) \equiv
(\tilde { r}_\mu(\tilde \tau))
= {\bf \tilde {\v R}}   + \delta {\bf {\tilde {\v r}}}(\tilde \tau)$. 
[Greek subscripts $\mu,\nu,\lambda$ are used to denote a collective index for a coordinate  and a particle index:  $\mu= (x,i)$ or $(y,i)$.]
Noting that $S[{\bf \tilde {\v r }}(\tilde \tau ) =  {\bf \tilde {\v R}} ]=0$,
\begin{fleqn}
    \begin{align}
    \label{eq:0 instanton}
    &
    \langle  \v R , \spin | e^{-\Delta \tau H} | \v R , \spin\rangle  \Big |_{\text{0-inst}}\! \sim  \left [ \det \! \Big ( \! \sqrt{r_s} \, {\bf  M}\big  [{\bf \tilde {\v r}}(\tilde \tau) =  {\bf \tilde {\v R}}  \big  ] \Big ) \! \right   ]^{ -  \frac 1 2}  
    \nonumber 
    \\
    &\! \equiv \!\! \int_{\delta {\bf {\tilde {\v r}}}(0)=  0}^{\delta {\bf {\tilde {\v r}}}(\Delta \tilde \tau )= 0} \!\!\!\!\!\!\!\!\! D \delta {\bf {\tilde {\v r}}}(\tilde \tau ) 
    \exp \!\left [ {  -\frac{ \sqrt{r_s} }{2} \!\!\! \int \!\! d\tilde \tau    
    \delta {\bf {\tilde {\v r}}}(\tilde \tau)^{\!T}  \v M[{\bf \tilde {\v R}} ]  
    \delta {\bf {\tilde {\v r}}}(\tilde \tau)
    } \right ]\!\! ,\!\!
     \\
    \label{eq:M definition}
    & M_{\mu\nu}  [ {\bf {\tilde {\v r}}}(\tilde \tau)   ] 
     \equiv 
     \frac{\delta^2 \tilde S}{\delta \tilde r_\mu(\tilde\tau) \, \delta \tilde r_\nu(\tilde\tau)} 
    =  -\delta_{\mu\nu}\frac{d^2}{d \tilde \tau^2}  
    +  \frac{\partial^2 V}{\partial \tilde r_\mu\partial \tilde r_\nu} [{{\bf {\tilde {\v r}}}(\tilde \tau) } ]
    \nonumber 
    \\
    & \ +\frac{2i}{\nu \sqrt{r_s}} 
    \left ( \mathcal E _{\mu \lambda} \dot {\tilde {r}}_\lambda  (\tilde \tau)  \partial_{\tilde r_\nu  }\tilde B [{\bf {\tilde {\v r}}}(\tilde \tau) ]
    +
    \mathcal E _{\mu\nu} \tilde B [{\bf {\tilde {\v r}}} ( \tilde \tau) ] \frac d {d\tilde \tau}
    \right ) 
    , 
    \\
    & 
     \mathcal E  \equiv \begin{bmatrix}
    0 & 1 & & & & \\
    -1 & 0 & & & & \\
    & & \ddots & & & \\
    & & & 0 & 1 \\
    & & & -1 & 0
    \end{bmatrix}.
    \label{eq:epsilon symbol}
    \end{align}
\end{fleqn}
When the  field is uniform $B( {\v r} ) =  B _0$, $\v M$ simplifies to 
\begin{fleqn}
    \begin{equation}
    M_{\mu\nu}  [ {\bf {\tilde {\v r}}} (\tilde \tau)   ] 
    \!=\!  -\delta_{\mu\nu}\frac{d^2}{d \tilde \tau^2} \!  
    + \! \frac{2i}{\nu \!  \sqrt{r_s}} 
    \mathcal E _{\mu\nu}  \frac d {d\tilde \tau}\!
    + \!  {\partial_{ \tilde r_\mu} \! \partial _{\tilde r_\nu} } \! V [{{\bf {\tilde {\v r}}}(\tilde \tau) } ].
    \end{equation}
\end{fleqn}
The numerator of \eqref{eq:exchange constant intermediate} is contributed by the semiclassical path together with harmonic fluctuations about it: ${\bf {\tilde {\v r}}}(\tilde \tau) = {\bf {\tilde {\v r}}}_a(\tilde \tau) + \delta {\bf {\tilde {\v r}}}(\tilde \tau)$.
Here, ${\bf {\tilde {\v r}}}_a(\tilde \tau)  $ is the solution to the saddle-point equation  of the dimensionless action \eqref{eq:action}
\begin{equation}
    \label{eq:equation of motion}
    \ddot{\bf {\tilde {\v r}}}_i -i \frac{2}{\nu \sqrt{r_s}} \dot {\bf {\tilde {\v r}}}_{i} \times  {\bf \tilde{\v  B}}[{\bf {\tilde {\v r}}}_{i} (\tilde \tau )]  
    - \nabla _{{\bf {\tilde {\v r}}}_i} V[{\bf {\tilde {\v r}}} (\tilde \tau)] =0,
\end{equation}
where ${\bf \tilde{\v  B}}({\bf {\tilde {\v r}}}) = \tilde{  B}({\bf {\tilde {\v r}}})   \hat z $,
satisfying the boundary condition 
\begin{equation}
    \label{eq:boundary condition}
    {\bf \tilde{\v r}}_a(0)  =   {\bf \tilde {\v R}}\equiv \v R  /r_0,   \   {\bf  \tilde{\v r}}_a (\Delta \tilde  \tau)  =  P_a {\bf \tilde { \v R}} \equiv  P_a\v R  /r_0.
\end{equation}
When $B\neq 0 $ (i.e.,  $1/\nu \neq 0$), no real solution to the above instanton equation exists; instead, ${\bf {\tilde {\v r}}}_a(\tilde \tau)$ is a {\it complex instanton} saddle of the path integral.
Defining $\tilde S_a \equiv \tilde S[{\bf {\tilde {\v r}}}_a]$, 
\begin{fleqn}
    \begin{align}
    \label{eq:1 instanton}
    &\langle P_a  (\v R , \spin) | e^{-\Delta \tau H} |  \v R , \spin \rangle\vert_{\text{1-inst}} \sim e^{-\sqrt{r_s} \tilde S_a} \times 
    \nonumber \\*
    &  \int_{\delta {\bf {\tilde {\v r}}}(0)= {\bf \tilde {\v R}}  } ^{\delta {\bf {\tilde {\v r}}}(\Delta \tilde \tau )= P_a {\bf \tilde {\v R}} } \!\!\!\! \!\!\!\!\!\!
    D \delta {\bf {\tilde {\v r}}}(\tilde\tau) \exp \! \left (\! {-\frac{ \sqrt{r_s} }{2} \!\! \int_0^{\Delta \tilde\tau}
    \!\!\!\! d\tilde \tau
    \delta {\bf {\tilde {\v r}}}(\tilde\tau)^T {\bf {M}} [ {\bf {\tilde {\v r}}}_a(\tilde\tau) ] \delta {\bf {\tilde {\v r}}}(\tilde\tau)}\! \right )
    \nonumber \\*
    &= e^{-\sqrt{r_s}\tilde S_a} \left [ \det\big (\sqrt{r_s} \, {\bf {M}} [ {\bf {\tilde {\v r}}}_a(\tilde\tau) ] \big ) \right ]^{-1/2}.
    \end{align}
\end{fleqn}
However, this integral  diverges due to the time-translational zero mode of  ${\bf {M}} [ {\bf {\tilde {\v r}}}_a(\tilde\tau) ]$: ${\bf {M}} [{\bf {\tilde {\v r}}}_a(\tilde\tau) ] \partial_{\tilde \tau}{\bf {\tilde {\v r}}}_a
(\tilde \tau) =0$.
After properly regularizing this divergence  \cite{kim2023dynamical, coleman1988aspects, zinnJustinQFT}, 
\begin{fleqn}
    \begin{align}
    \label{eq:1 instanton zero mode}
    &\langle  P_a  (\v R , \spin) | e^{-\Delta \tau H} |  \v R , \spin  \rangle\vert_{\text{1-inst}
    } 
    \\*
    &=
    \Delta \tau E_{\rm Debye}
    \sqrt{\frac{\tilde I_a}{2\pi}}   e^{-\sqrt{r_s}\tilde S_a} \big [ {\rm det}' \big ( \sqrt{r_s} \, {\bf {M}} [{\bf {\tilde {\v r}}}_a(\tilde\tau) ] \big ) \big ]^{-\frac 1 2}, \nonumber
    \\
    &
    \label{eq:I_a}
    \tilde I _ a \equiv 
    \int_0^{\Delta \tilde \tau}       \dot{\bf {\tilde {\v r}}}_a  ^2 \   d\tilde\tau
    =
    \int_0^{\Delta \tilde \tau}  d\tilde\tau  \left (  \frac 1 2 {\dot{\bf {\tilde {\v r}}}_a^2 }  
    + V ({\bf {\tilde {\v r}}}_a ) - \tilde V_0 \right ),
    \end{align}
\end{fleqn}
where $\tilde I _ a $ is the action \eqref{eq:action} without the magnetic field contribution and  $\det '$ is the  determinant  without the zero mode.
The second equality in \eqref{eq:I_a} follows from the Euclidean ``energy conservation'' implied by  \eqref{eq:equation of motion}
\begin{equation}
    \label{eq:conservation of energy}
     \frac 1 2 \left [{\dot{\bf {\tilde {\v r}}} _a (\tilde \tau ) } \right ] ^2  -  V[{\bf {\tilde {\v r}}}_a(\tilde \tau)]  + \tilde V_0 = 0.
\end{equation}
Using the identity, 
\begin{fleqn}
    \begin{equation}
    \left (\frac{  {\rm det}' \big ( \sqrt{r_s}  {\bf {M}} [ {\bf {\tilde {\v r}}}_a(\tilde\tau) ] \big )}{ {\rm det} \big ( \sqrt{r_s}  {\bf {M}} [ {\bf \tilde {\v R }}  ] \big )} \right )^{ \!\!-\frac 1 2}
    \!\!\!\!
    =\!
    r_s^{1 /4}   \!\left (  \frac{  {\rm det}' \,    {\bf {M}} [ {\bf {\tilde {\v r}}}_a(\tilde\tau) ] }{ {\rm det}  \,   {\bf {M}} [ {\bf \tilde {\v R }}   ]  } \right )^{\! -\frac 1 2}\!\!\!\! ,\!
    \end{equation}
\end{fleqn}
we finally obtain
\begin{align}
    \label{eq:exchange constant final}
    J_a 
    &=
    \frac{e^2}{4\pi\epsilon a_{\rm B}}
    \cdot \frac {{A_a}} {{ r_s^{5/4} }}\sqrt{\frac{\tilde I_a}{2\pi}}\ e^{-\sqrt{r_s}\tilde S_a},
     \\*
     \label{eq:fluctuation determinant}
    A_a &= 
        \left ( \frac{ {\rm det}' \,  \v {\bf {M}} [ {\bf {\tilde {\v r}}}_a(\tilde\tau) ] }{\det\,  {\bf {M}} [{\bf  \tilde {\v R}}  ]} \right )^{\!\! -\frac 1 2},
\end{align}
where $\v M [{\bf \tilde {\v r}} (\tilde \tau)  ]$ and $\tilde I _a $ are given by  \eqref{eq:M definition} and \eqref{eq:I_a}, respectively, and $\tilde S_a \equiv \tilde S[{\bf \tilde {\v r}}_a  ]$.
The effective Hamiltonian \eqref{eq:WC exchanges} is obtained by summing over all instanton processes $P_a$
\footnote{
    Summing over instanton configurations with multiple distinct instanton events $a$ in the zero-temperature limit ($\beta \gg 1/|J_a| $) requires one to first  divide  $\tilde \beta = \beta E_{\rm Debye} $ into $M$ time slices, $\tilde \beta = M \Delta \tilde \tau$, where  $\Delta \tilde {\tau }$ is at intermediate time scale \eqref{eq:intermediate time scale} and then take $M\to \infty$ limit. 
    For a detailed treatment of this subtle derivation, we refer readers to  \cite{voelker2001disorder}.
}.

\ksk{
\subsection{Small Cyclotron Condition $  ( {\nu \sqrt{r_s }})^{-1} \ll  1 $}}

The expression \eqref{eq:exchange constant final} is valid for any value of $B$ [in particular, for a large field such that $\nu \sqrt{r_s } = O(1)$], provided  the complex instanton solutions of (\ref{eq:equation of motion}-\ref{eq:boundary condition}) and the resulting fluctuation determinant (\ref{eq:M definition}, \ref{eq:fluctuation determinant}) are obtained (which is a complex task). 
But since the $B=0$ real instanton solutions [$\equiv {\bf {\tilde {\v r}}}_a^{(0)}(\tilde \tau)$] are already known  \cite{voelker2001disorder, Katano2000WKB, kim2023dynamical}, we would like to obtain a simplified expression for $J_a$ by expanding ${\bf {\tilde {\v r}}}_a(\tilde \tau)$ around  ${\bf {\tilde {\v r}}}_a^{(0)}(\tilde \tau)$ in powers of  $1/\sqrt r_s$:
\begin{align}
    {\bf {\tilde {\v r}}}_a(\tilde \tau) = {\bf {\tilde {\v r}}}_a^{(0)}(\tilde \tau)  + \frac{1}{\sqrt r_s} \delta {\bf {\tilde {\v r}}}_a^{(1)}(\tilde \tau) + 
    O\left (  r_s^{-1} \right).
\end{align}
\ksk{Formally, this asymptotic expansion is valid when the perturbation---namely, the second term in \eqref{eq:equation of motion}---is small compared to the unperturbed terms, i.e., when $(\nu\sqrt r_s)^{-1} \ll 1$.
We refer to this as the ``small cyclotron condition'' since it is equivalent to requiring that the cyclotron energy scale $(\hbar \omega_c =\hbar eB_0/m)$ is much smaller than the typical oscillator energy scale ($\sim E_{\rm Debye}$) [See Footnote \cite{footnote:small_parameter} for more detail].
In this work, we always consider the semiclassical $r_s\to \infty$ limit and work at a fixed Landau-level filling $\nu = O(1)$, so that the weak-field condition is automatically satisfied.} Substituting this in \eqref{eq:equation of motion}  results in,  at zeroth order, a zero-field real instanton equation
\begin{equation}
    \label{eq:real instanton equation}
    \ddot {\bf {\tilde {\v r}}}_a^{(0)}  -  \nabla_{{\bf {\tilde {\v r}}}} V [{\bf {\tilde {\v r}}}_a^{(0)}(\tilde \tau) ]= 0,
\end{equation}
subject to the boundary condition \eqref{eq:boundary condition}.
At first  order,  
\begin{fleqn}
    \begin{equation}
    \label{eq:1st order}
      \delta \ddot {\bf {\tilde {\v r}}}_{i,a}^{(1)}
        - \frac{2i}{\nu} \dot {\bf {\tilde {\v r}}}_{i,a}^{(0)} \times {\bf   \tilde {\v B}}[\tilde {\v r}_{i,a}^{(0)}] 
        - \! \left ( \nabla_{{\bf {\tilde {\v r}}}_i}\! \nabla_{ {\bf {\tilde {\v r}}}_j } \! V  [{\bf {\tilde {\v r}}}_{a}^{(0)} ] \right ) \delta {\bf {\tilde {\v r}}}_{j,a}^{(1)} = 0,\!
    \end{equation}
\end{fleqn}
with $\delta {\bf {\tilde {\v r}}}^{(1)}_a(0) = \delta {\bf {\tilde {\v r}}}^{(1)}_a(\Delta \tilde \tau) =  0$, where the index $j$ is implicitly summed over.
Since this is a second-order linear inhomogeneous differential equation for $\delta {\bf {\tilde {\v r}}}^{(1)}_a$ with imaginary functional coefficients, $\delta {\bf {\tilde {\v r}}}^{(1)}_a$ is purely imaginary. 
The solution ${\bf {\tilde {\v r}}}_a^{(0)} $ to Eq. (\ref{eq:real instanton equation}) for the $J_3$ process is illustrated in Fig. \ref{fig:tunneling} (a).

Plugging this solution into the action \eqref{eq:action}, we obtain
\begin{fleqn}
    \begin{align}
    \label{eq:perturbative instanton action}
       & \tilde S_a 
          = \int_0 ^ {\Delta \tilde \tau}  d\tilde \tau   \left [  \dot{\bf {\tilde {\v r}}}_a ^2   +  \frac{2i}{\nu\sqrt{r_s} }  \, \dot {\bf {\tilde {\v r}}}_{a}   \cdot  {\bf  \tilde {\v A}} ({\bf {\tilde {\v r}}}_a )  \right ] 
          \nonumber \\
           &  \ \ \ 
           = \tilde S_a^{(0) }  -  \frac i {\sqrt r_s } \phi_a [B(\v r )] + O\left( 1/  { r_s } \right ) 
        \\
        \label{eq:real instanton action}
         & \tilde S_a^{(0)} =  \int_0 ^ {\Delta \tilde \tau} d\tilde \tau   \left ( \dot{\bf {\tilde {\v r}}}_a^{(0)} \right )  ^2 = \int d \tilde r _a^{(0)}\sqrt{2 V[{\bf {\tilde {\v r}}}_a^{(0)} ]}, 
            \\
        \label{eq:Aharonov-Bohm-dimensionless}
         &\phi_a  [B (\v r)]
         \!=\! - \frac 2 \nu  \! \int  \! d {\bf {\tilde r}}^{(0)}_{a} \!\cdot\! {\bf \tilde {\v A}} ({\bf {\tilde r}}^{(0)}_{a})
         \!=\! - \frac{e}{\hbar}   \!\int\! d\v r^{(0)}_{a} \!\cdot \! { \v A} (\v r^{(0)}_{a}),\!\!\!
    \end{align}
\end{fleqn}
where  $\tilde S^{(0)}_a \! \equiv  \tilde S[\v r ^{(0)}_a] > \! 0 $ is the zero-field instanton action   
and $\v A(\v r_a) 
\!\equiv\!  ( \v A(\v r_{1,a} ), \cdots ,\v A(\v r_{N,a} ))$.
The first line of (\ref{eq:perturbative instanton action}) follows from the Euclidean energy conservation  \eqref{eq:conservation of energy}, and the second line from the following  identity:
\begin{align}
    \label{eq:zero off-diagonal}
    &\int d\tilde \tau \dot{\bf {\tilde {\v r}}}_a^{(0)}  \cdot  \delta \dot{\bf {\tilde {\v r}}}_a^{(1)}  = \int d\tilde \tau \nabla_{\bf {\tilde {\v r}}} V[{\bf {\tilde {\v r}}}_a^{(0)} ] \cdot \delta \dot{\bf {\tilde {\v r}}}_a^{(1)}  
    \nonumber \\
    &= 
    \int d\tilde \tau \ddot{\bf {\tilde {\v r}}}_a^{(0)} \cdot  \delta {\bf {\tilde {\v r}}}_a^{(1)} = -\int d\tilde \tau \dot{\bf {\tilde {\v r}}}_a^{(0)} \cdot  \delta \dot{\bf {\tilde {\v r}}}_a^{(1)}  = 0.
\end{align}
Here, the first equality is due to the first-order energy conservation that follows from \eqref{eq:conservation of energy},
\begin{align}
    \dot{\bf {\tilde {\v r}}}^{(0)} \cdot \delta \dot{\bf {\tilde {\v r}}}^{(1 )} -  \nabla_{\bf {\tilde {\v r}}} V[{\bf {\tilde {\v r}}}^{(0)}]\cdot \delta {\bf {\tilde {\v r}}}^{(1)} =0,
\end{align}
the second uses  \eqref{eq:real instanton equation} and the third is integration by parts. 
The expression \eqref{eq:zero off-diagonal} is $0$ since the first and last expression are equal up to a minus sign. 
For a uniform field, $B(\v r ) = B_0 $, we obtain a simpler expression (\ref{eq:Aharonov-Bohm_uniform}-\ref{eq:area})
\begin{align}
    \label{eq:main-Aharonov-Bohm-uniform}
        & \phi_a [B_0] = - \frac{2}{\nu}  \sum_i  \frac 1 2 \int {\bf {\tilde r}}_{i,a
    }^{(0)} \times  d {\bf {\tilde r}}_{i,a}^{(0)}  = 
    - \frac{2}{\nu}\tilde \Sigma _a ^{(r) },
\end{align}
where  
$\tilde \Sigma_a$ is the signed area (positive for a counter-clockwise motion) enclosed by all ${\bf \tilde {\v r }}_{i,a}^{(0)}(\tilde \tau)$.

The imaginary deformation of the instanton path results in an $O( r_s^{-1/2})$ correction  in the fluctuation determinant \eqref{eq:fluctuation determinant}  relative to its zero-field value $A_a^{(0)}$
\begin{align}
  &  A_a = A_a^{(0)} + O\left ( 1/\sqrt r_s \right ),
    \\
    \label{eq:fluctuation determinant zeroth-order}
   & A_a^{(0)} \equiv   \left [ \frac{ {\rm det}'\big (  -\delta_{\mu\nu}\frac{\partial^2}{\partial \tilde \tau^2}  +  \frac{\partial^2V}{\partial \tilde r_\mu\partial \tilde r_\nu} \big [{\bf {\tilde {\v r}}}^{(0)}_a(\tilde\tau) \big ]\big )}{\det\big (  -\delta_{\mu\nu}\frac{\partial^2}{\partial \tilde \tau^2} +  \frac{\partial^2V}{\partial \tilde r_\mu\partial \tilde r_\nu} [ {\bf \tilde {\v R}}  ] \big )} \right ]^{-\frac 1 2}. 
\end{align}
Therefore, to leading order,  \eqref{eq:exchange constant final} reduces to (\ref{eq:exchange1}-\ref{eq:Aharonov-Bohm}):
\begin{fleqn}
    \begin{align}
    \label{eq:main-Aharonov-Bohm}
    J_a[B(\v r )]   &= 
    \frac{e^2}{4\pi\epsilon a_{\rm B}} \cdot 
     \frac {{A_a^{(0)}}} {{ r_s^{5/4} }}\sqrt{\frac{\tilde S_a^{(0)}}{2\pi}}\ e^{-\sqrt{r_s}\tilde  S_a^{(0)}}  e^{i \phi_a{\left [ B (\v r) \right ] }  }. \! 
    \end{align}
\end{fleqn}
For a uniform magnetic field, $B[\v r] =  B _0$, the Aharonov-Bohm phase $\phi_a[B _0]$ depends only on  $\nu$ and the dimensionless area  $\tilde \Sigma_a^{(r)}$ 
\eqref{eq:main-Aharonov-Bohm-uniform}.
The numerically calculated values of $\tilde S_a^{(0)}$, $A_a^{(0)}$ and $\Sigma_a^{(r)}$ are  reported in Table \ref{table:results}.
\\

\section{Effects of a Berry curvature}

Here, we generalize the the previous results and analyze the combined effects of a Berry curvature,  $\Omega (\v k) =  \nabla_{\v k} \times \A (\v k)  = \Omega_0 + \delta \Omega (\v k)$, and a magnetic field.
For simplicity, the magnetic field is assumed to be uniform, $B(\v r) = B_0  = \nabla _{\v r } \times \v A(\v r)$.
The real-time dynamics of the 2DEG in this case is captured by the phase-space action \cite{xiao2010berry, chang1996berry}
\footnote{
    To faithfully capture the Berry curvature effect within the Hamiltonian approach (instead of the Lagrangian approach), one needs to include at least two bands as in \cite{tan2024parent, soejima2025jellium}. 
    This is not necessary for our purpose as we work with the effective continuum theory.
}
\begin{align}
\label{eq:Minkowski}
    S_M   & = \int dt \Bigg ( 
    \sum_i \bigg  [ \hbar \v k_i  \cdot  \frac{d {\v r }_i}{dt} -  \varepsilon_M(\v k _i) 
    - e \frac{d \v r _i }{dt}  \cdot \! \v  A (\v r_i)
    \nonumber \\
    &  \ \ \ 
    +  \frac{d \v k _i }{dt}  \cdot  \A  (\v k_i) \bigg ] - \frac{e^2}{4\pi \epsilon}  \sum_{i<j }  \frac{1}{|\v r_i - \v r_j|} + V_0  \Bigg ),
\end{align}
where, as in \eqref{eq:action-unnormalized}, the Coulomb energy is measured relative to its  minimum  value $V_0 = \frac {e^2 }{4\pi \epsilon r_0 } \tilde  V_0 $. 
\begin{equation}
\label{eq:effective mass approximation}
    \!\!\!\!\!\!
    \varepsilon_{\!M}(\v k )  \!=\! \frac {\hbar^2 k^2 }{2m}  \!- \! B_0 M_z(\v k + \v Q_0 ) 
    \!\approx\! \frac {\hbar^2 k^2 }{2m^*}  \!-\! B_0 M_z(\v Q_0)\!
\end{equation}
is the  single-electron energy including  the coupling of  the  orbital magnetic moment $M_z$ to the magnetic field, where $\v Q_0$  is the momentum at the band minimum.
The continuum approximation near $\v k \sim 0$ merely results in the effective-mass renormalization $m\to m^*$  \eqref{eq:effective band mass} with a constant shift in energy 
\footnote{
    The details of this effective-mass renormalization depends on the parent band structure from which our effective model is assumed to be derived.
    We do not explore such model-dependent details here, but instead treat $m^*$ (and $M_z$) as a given parameter.
}.
This in turn renormalizes the effective Bohr radius 
$a_B ^* =  4\pi \epsilon \hbar^2 /(m^* e^2)$, the $r_s$-parameter $r_s^* = r_0/a_B^*$ and the Hartree energy $E_h^* = e^2/(4\pi \epsilon a_B^* )$.

\ksk{In this work, we primarily focus on the spin physics in the case where only a single valley near $\v Q_0$ is active at low energy. In the presence of the time-reversal symmetry (TRS), valleys come in TR-related pairs at $\pm \v Q _0$ in the non-interacting band structure, with opposite Berry curvature (and opposite orbital magnetic moment). We therefore assume that the valley is fully polarized due to interactions and the TRS is spontaneously broken, so that the single-valley description with $\Omega \neq 0$ is appropriate. 
Within this projected single-valley theory, the term $- B_0 M_z(\v Q_0)$ is constant and can be dropped without affecting the spin physics 
\footnote{
    If multiple valleys with different orbital magnetic moments are active at low energy, then $- B_0 M_z(\v Q_0)$ acts as a Zeeman-like term for valley degrees of freedom and must be retained.
    Relatedly, a trigonal-warping correction to the otherwise isotropic dispersion $\v p^2/2m^*$ can drive  spontaneous valley polarization of the WC even without the Zeeman-like term in \eqref{eq:effective mass approximation} \cite{calvera2022WC}.
}.}

Since we are interested in  thermodynamic properties, we work with the imaginary-time Euclidean action ($\tau = it/\hbar$)
\begin{fleqn}
    \begin{align}
        \label{eq:Euclidean}
        & S_E =\!\int d\tau  \Bigg ( 
        \sum_i \bigg [ -i  \v k_i \! \cdot  \frac {d{\v r }_i}{d\tau } + \frac {\hbar^2 k_i^2 }{2m^*} + i  e \frac{d \v r _i }{d\tau}  \cdot  \v  A (\v r_i) 
        \nonumber \\
        & \!
        -i    \frac{d \v k _i }{d\tau} \! \cdot \! \A  (\v k_i) \bigg ]
         +\frac{e^2 }{4\pi \epsilon }  \sum_{i<j} \frac 1 {|\v r_i - \v r_j|} -  V_0 
        \!\Bigg ) \! 
        \equiv   \sqrt {r_s^*} \tilde S,\!
    \\
        \label{eq:action for Berry curvature}
        & \tilde S  
        =\int d \tilde \tau  \Bigg ( 
        \sum_i \bigg [ -i  {\bf \tilde {\v k}}_i  \cdot  \dot{\bf {\tilde {\v r}}}_i +  
        \frac {1 }{2} \tilde  k_i^2
     +   \frac {2 i} {\nu \sqrt {r_s^*} }  \dot {\bf {\tilde {\v r}}}_{i}   \cdot {\bf \tilde {\v A }} ({\bf {\tilde {\v r}}}_i)
    \nonumber \\
    &  \ \ \ \ \ \ \ \ \ \ \ \ \ \ \ \ \ \ \ \ 
    - \frac{i \alpha^*}{2 \sqrt {r_s^* }} 
    \dot {\bf \tilde {\v k}}_{i}  \cdot {\bf \tilde {\A}} ({\bf \tilde {\v k}}_i)
    \bigg ]
    + V({\bf {\tilde {\v r}}}) -\tilde V_0 \Bigg ),
    \\
    &
    \label{eq:dimensionless Berry curvature}
    \boldsymbol{ \tilde {\mathcal A }}({\bf \tilde {\v k}}) \!=\! \frac {\!\sqrt {a_B^* r_0}} {\Omega_0} \! \boldsymbol{  {\mathcal A }}(\v k),
    \, \tilde \Omega ({\bf \tilde {\v k}}) \!=\! \nabla_{\bf \tilde {\v k}} \!\!\times\! \boldsymbol{\tilde {\mathcal A}}({\bf \tilde {\v k}})
    \!= 
    \! 1 \!+\! \frac {\delta \Omega({ {\v k}} )}  {\Omega _0} .\!\!
    \end{align}
\end{fleqn}
Again, $r_s^*$ dependence of the action becomes manifest when expressed in terms of   dimensionless coordinates and momenta: ${\bf {\tilde {\v r}}} \equiv {\v r}/r_0$, $\tilde \tau = \tau E_{\rm Debye}$ and  ${\bf \tilde {\v k}} \equiv \sqrt{a_B ^* r_0}\  \v k  = r_0 \v k / \sqrt {r_s^*}$.
$\boldsymbol {\tilde {\mathcal A}}({\bf \tilde {\v k}}) $  and $\tilde \Omega  ({\bf \tilde {\v k}})$ are the dimensionless Berry connection and Berry curvature, respectively.
$\alpha^* \equiv  r_s^*  \nu_{\Omega} =  2\Omega_0/(a_B^{*2} r_s^*)$   is a parameter  that captures the Berry phase effect \eqref{eq:alpha} and is assumed to be $O(1)$. $\nu_{\Omega} \equiv \Omega_0 \cdot 2\pi n $  \eqref{eq:Chern filling factor}.
Since the magnetic field is assumed to be uniform, the dimensionless magnetic field has a unit magnitude ${\tilde{B}} =  1 $.

\ksk{
\subsection{Semiclassical Limit and Small Berry-Confinement Condition $\alpha/\sqrt r_s \ll 1$}}
With this setup, the exchange constants $J_a[B,\Omega]$ can be derived analogously from the large-$r_s$ expansion with some  modifications.
The saddle-point equations of the action \eqref{eq:action for Berry curvature} are now  first-order Hamilton's equations for complex phase-space variables $({\bf \tilde {\v r}}_i, {\bf \tilde  {\v k}}_i)$, $i=1,\cdots,N$, 
\footnote{
    The Hamilton's equations (\ref{eq:Hamilton equation for r}-\ref{eq:Hamilton equation for k}) do not, in general, decouple nicely in terms of either $\v r $ or $\v k$.
    However, when $\Omega(\v k) = \Omega_0 $ is constant, they can be combined to yield
    \begin{align*}
        \ \ \ \ \ \left ( 1 + \frac {\alpha^*} {\nu r_s^*} \! \right ) \!\ddot{\bf \tilde {\v r}}
        \!-\!
         \frac {2i }{\nu \sqrt {r_s^*}} \mathcal E   \dot{\bf \tilde {\v r}} 
        \!+\! \frac {i \alpha^* }{2 \sqrt {r_s^*}} \mathcal E \, \nabla_{\!{\bf \tilde {\v r}}}\! \nabla_{\!{\bf \tilde {\v r}}} V[{\bf \tilde {\v r}}]    \, \dot{\bf \tilde {\v r}} 
        -\! \nabla_{\!{\bf \tilde {\v r}}} V[{\bf \tilde {\v r}}] = 0,
    \end{align*}
    where $\nabla_{{\bf \tilde {\v r}}}\nabla_{{\bf \tilde {\v r}}} V[{\bf \tilde {\v r}}]$ is the Hessian matrix of $V$ at ${\bf \tilde {\v r}}$.
}:
\begin{align}
    \label{eq:Hamilton equation for r}
    \dot {\bf \tilde {\v r }}_i &= - i  {\bf \tilde {\v k }}_i - \frac {\alpha^*} {2 \sqrt {r_s ^*}}  \dot {\bf \tilde {\v k }}_i \times  {\bf \tilde \Omega} ({\bf \tilde {\v k}}_i),
    \\
    \label{eq:Hamilton equation for k}
    \dot {\bf \tilde {\v k }}_i & =  i  \nabla_{\bf \tilde {\v r}_i} V - \frac 2 {\nu  \sqrt {r_s^*}}  \dot {\bf \tilde {\v r  }}_i \times   \bf {\tilde {B} },  
\end{align}
where ${\bf \tilde \Omega}({\bf \tilde {\v k}}) \equiv {\tilde \Omega}({\bf \tilde {\v k}})\hat z$ and ${\bf \tilde {\v B}}({\bf \tilde {\v r }}) = \hat z $. \ksk{These are the imaginary-time version of the well-known semiclassical equations of motion \cite{xiao2010berry}.}
We solve these equations  perturbatively in  $1/\sqrt {r_s^*}$:
\begin{align}
     {\bf \tilde{\v r}}_a(\tilde \tau) &= {\bf \tilde{\v r}}_a^{(0)}(\tilde \tau)  + \frac{1}{\sqrt {r_s^*}} \delta {\bf \tilde{\v r}}_a^{(1)}(\tilde \tau) + 
    O\left (  1 / {r_s^*} \right),
    \\
    {\bf \tilde{\v k}}_a(\tilde \tau) &= {\bf  \tilde{\v k}}_a^{(0)}(\tilde \tau)  + \frac{1}{\sqrt {r_s^*}} \delta {\bf \tilde{\v k}}_a^{(1)}(\tilde \tau) + 
    O\left ( 1 /  {r_s^*} \right).
\end{align}
\ksk{Again, this asymptotic expansion is valid when the last terms of (\ref{eq:Hamilton equation for r}-\ref{eq:Hamilton equation for k}) are small compared to the other terms, i.e., when $(\nu \sqrt {r_s^*})^{-1} \ll 1$ and  $\alpha^*  /\sqrt {r_s ^*}\ll 1$. The former corresponds to the small cyclotron condition \cite{footnote:small_parameter}, while the latter defines the analogous ``small Berry-confinement condition,'' where the characteristic Berry-confinement energy scale 
\begin{align}
    E_{\Omega} \equiv \frac 1 2 \left ( \frac {e^2} {2\pi \epsilon r_0^3 } \right ) \ell _{\Omega_0}^2  = E_h \cdot \frac {\alpha}{2r_s^2}
\end{align}
is much smaller than the Debye energy scale $E_{\rm Debye}$ 
\footnote{
\ksk{Here, $E_{\Omega}$ could be understood as the quantum zero-point energy scale associated with the harmonic oscillator $V= \frac 1 2 K_{\rm typ}(X^2 + Y^2)$, where $K_{\rm typ} \sim \frac {e^2} {2\pi \epsilon r_0^3} = \frac {d^2} {dr^2 } \frac{e^2}{4\pi \epsilon r}\Big |_{r=r_0} $ is the appropriate ``spring constant'' for the Coulomb interaction. Since we work with the effective Lagrangian within a single band, the band-projected position operators  do not commute due to the presence of a Berry curvature, $[X,Y] \approx i \Omega(\v k)$ \cite{xiao2010berry,parameswaran2012fractional, marzari1997maximally}. Hence, $V$ can be interpreted as a harmonic oscillator with $Y$ being the  momentum canonically conjugate to $X,$ giving rise to the associated Berry-confinement energy scale $E_{\Omega}$.}}.
Since we fix $\nu,\alpha^* =O(1)$ and take the semiclassical limit $r_s^* \to \infty$, both conditions are automatically satisfied.}
The zeroth-order equations in this expansion are
\begin{equation}
    \label{eq:Hamilton's eq at 0th order}
      \dot{\bf {\tilde {\v r}}}_{i,a}^{(0)}
     =-i {\bf \tilde {\v k}}_{i, a}^{(0)} 
     \ \text{ and  }\ 
     \dot{\bf  {\tilde {\v k}}}_{i, a}^{(0)}
     = i \nabla_{{\bf {\tilde {\v r}}}_i } V [{\bf {\tilde {\v r}}}_a^{(0)} ] 
\end{equation}
subject to the boundary condition 
${\bf {\tilde {\v r}}}_a^{(0)}(0) \! =\! {\bf \tilde {\v R}}  \! = \! { \v R} /r_0$ and ${\bf \tilde{\v r}}_a^{(0)} (\Delta \tilde  \tau) = P_a {\bf \tilde { \v R}}.$
The first-order equations are 
\begin{align}
\label{eq:Hamilton's eq for r at 1st order}
    \delta \dot  {\bf {\tilde {\v r}}}_{i,a}^{(1)} 
    &=  -i 
    \delta {\bf {\tilde k}}_{i,a}^{(1)} 
        - \frac {\alpha^*} 2
        \dot {\bf {\tilde {\v k}}}_{i,a}^{(0)} \times {\bf \tilde \Omega} ({\bf {\tilde k}}_{i,a}^{(0)}),
        \\
    \label{eq:Hamilton's eq for k at 1st order}
    \delta \dot  {\bf {\tilde {\v k}}}_{i,a}^{(1)} 
    &=  i  \nabla_{{\bf \tilde {\v r}} _{i}} \!\nabla_{{\bf  \tilde {\v r}} _{j}}\! V [{\bf {\tilde {\v r}}}_{a}^{(0)} ] \,  
    \delta {\bf {\tilde {\v r}}}_{j,a}^{(1)} 
    - \frac   2 \nu  \dot {\bf {\tilde {\v r}}}_{i,a}^{(0)} \times {\bf  \tilde {\v B}}         
\end{align}
with $\delta {\bf {\tilde r}}_a^{(1 )}(0)  =  \delta {\bf {\tilde r}}_a^{(1 )}(\Delta \tilde \tau) = 0$.
It is evident that ${\bf \tilde {\v r}}_a^{(0)}$ is real since it satisfies the real instanton equation \eqref{eq:real instanton equation} 
and hence ${\bf {\tilde k}}_a^{(0)} = i \dot {\bf {\tilde {\v r}}}_a^{(0)} $ is purely imaginary.
In contrast, $\delta {\bf {\tilde r}}_a^{(1)}$ and $\delta {\bf {\tilde k}}_a^{(1)}$ are in general complex-valued since $\tilde \Omega({\bf \tilde {\v k}}_a^{(0)})$ is complex for an imaginary argument
\footnote{
    For an isolated band, the Berry curvature is a smooth periodic function in a real momentum space and  has a Fourier  expansion. 
    The analytic continuation to the complex momentum follows  straightforwardly.
    In the special case where the Berry curvature is constant, $\Omega(\v k) = \Omega_ 0$, $\delta \tilde {\v r}_a^{(1)}$ becomes purely imaginary and  $\delta \tilde {\v k}_a^{(1)}$ purely real.
}.
The solution $({\bf {\tilde {\v r}}}_a^{(0)},{\bf {\tilde {\v k}}}_a^{(0)}) $ to Eq. (\ref{eq:Hamilton's eq at 0th order}) for the $J_3$ process is illustrated in Fig. \ref{fig:tunneling}.

With the solutions to (\ref{eq:Hamilton's eq at 0th order}-\ref{eq:Hamilton's eq for k at 1st order}), the tunneling action can be obtained straightforwardly with similar manipulations as in the  magnetic-field-only case:
\begin{fleqn}
    \begin{align}        \label{eq:perturbative instanton action for Berry}
    & \tilde S_a 
     =   \tilde S_a^{(0)}  -
    \frac i {\sqrt {r_s^*} } \left (
    \phi_a[B_0] + \gamma_a [\Omega(\v k)]
    \right ) 
    +  O\left (  1 /  {{r_s^*}}    \right),
    \\ 
    \label{eq:Berry-phase-main}
    & \gamma_a [\Omega(\v k)] = 
    \! \sum_i \! \oint d\v k^{(0)}_i \!\!\cdot\! \boldsymbol{\mathcal A}  (\v k^{(0)}_i) = \frac {\alpha^*} 2 \! \sum_i\! \oint d {\bf  {\tilde k}}^{(0)}_i \!\! \cdot \! \boldsymbol{\tilde {\A}}  ({\bf {\tilde k}}^{(0)}_i), 
    \end{align}
\end{fleqn}
where  $\tilde S_a^{(0)}$ is the real instanton action \eqref{eq:real instanton action} and  $\phi_a[B_0]$ is the Aharonov-Bohm phase \eqref{eq:main-Aharonov-Bohm-uniform}. 
For a uniform Berry curvature 
$\Omega(\v k) = \Omega_0,$
\begin{equation}    
    \label{eq:dimensionless area momentum}
    \gamma_a [\Omega_0 ] =  \frac {\alpha^*} 2    \sum_i  \frac 1 2 \oint {\bf {\tilde  k}}_{i,a
        }^{(0)} \times  d  {\bf {\tilde  k}}_{i,a}^{(0)}
        \equiv 
         \frac {\alpha^*} 2 \tilde \Sigma_a^{(k)}.
\end{equation}
The fluctuation determinant receives only a subleading $O(1/\sqrt  {r_s^*})$ correction which we ignore.
Therefore,
\begin{align}
    \label{eq:exchange_field and Berry}J_a[B_0,\Omega(\v k) ] = J_a^{(0)} (E_h^*, r_s^* )\, 
    e^{i \left ( \phi_a[B_0] +  \gamma_a [\Omega(\v k)]
    \right )},
\end{align}
with the magnitude given by \eqref{eq:exchange-constant-introduction} with properly renormalized Hartree energy  $E_h^*$ and the $r_s$-parameter $r_s^*$.
\ksk{Note that, when Berry curvature is absent $\Omega({\bf k}) =0$, \eqref{eq:exchange_field and Berry} reduces to the result of the previous section \eqref{eq:main-Aharonov-Bohm}.}

\section{Discussion}

We analyzed how a magnetic field $B$ and a Berry curvature $\Omega$ modify the exchange interactions in a Wigner crystal (WC) via an asymptotic large-$r_s$ expansion.
When either $B$ or $\Omega$ is present individually, the effect is simply to multiply the  exchange constant $J_a^{(0)}$ at $B=\Omega = 0$ by a phase factor: 
the Aharonov-Bohm phase $\phi_a(B)$ for $B\neq 0$ (\ref{eq:exchange1}-\ref{eq:area}) and the Berry phase $\gamma_a(\Omega) $ for  $\Omega \neq 0$ (\ref{eq:exchange2}-\ref{eq:area momentum}).
When both are  present, in addition to having both phases, the exchange magnitude $|J_a[B,\Omega]|= J_a^{(0)}(E_h^*,r_s^*)$ is also modified via an effective-mass renormalization  (\ref{eq:effective band mass}) originating from the coupling of the orbital magnetic moment to $B$.

\begin{figure}[t]
    \centering
    \includegraphics[width=.9\linewidth]{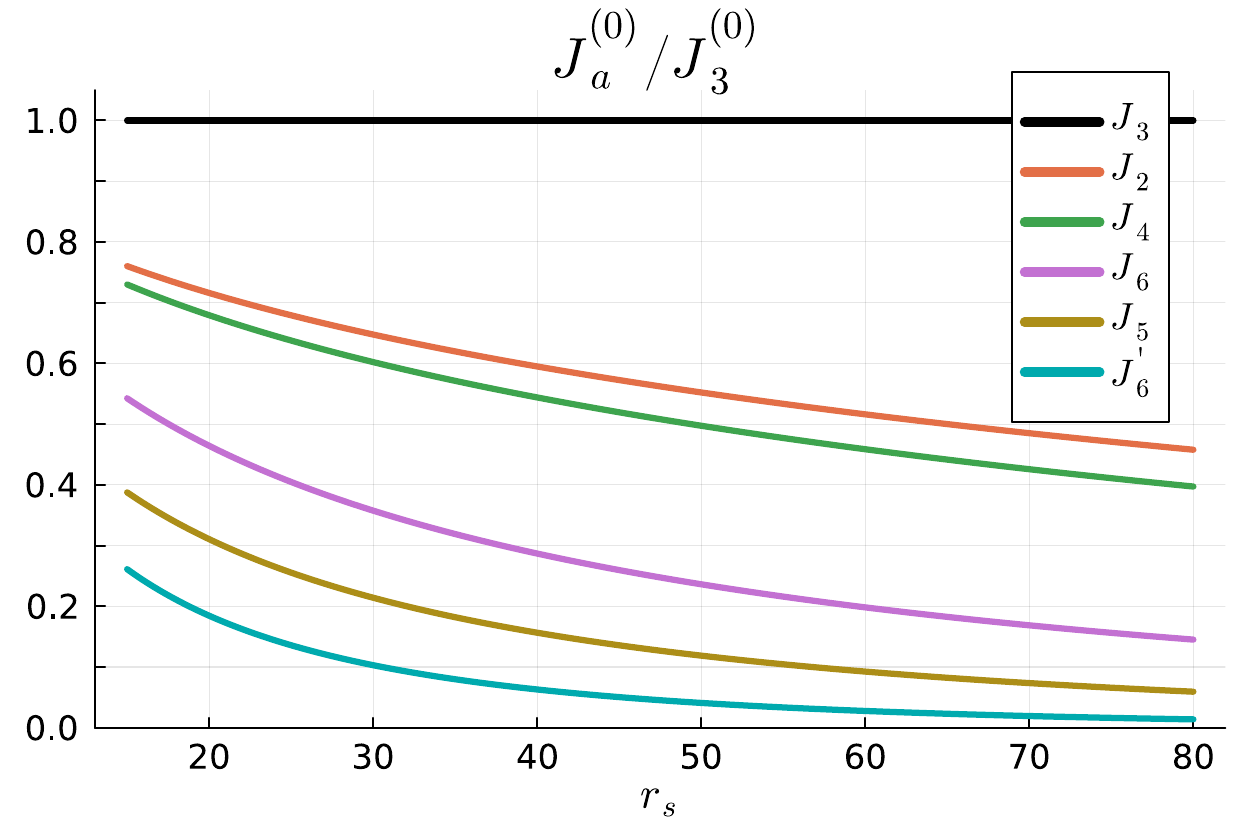}
    \caption{Exchange magnitude  $J_a^{(0)}$ [see \eqref{eq:exchange-constant-introduction} and Table \ref{table:results}] normalized by the largest one $J_3^{(0)}$ as a function of $r_s$ ($15 \leq r_s \leq 80$).
     }
     \label{fig:exchange ratio}
\end{figure}

In the absence of $B$ and $\Omega$, the WC is thought to occur for $r_s \gtrsim 37$ \cite{smith2024unified}, aside from the subtleties of possible intermediate phases \cite{footnote:intermediate-phase}.
(This critical $r_s$ value will generally shift when $B$ or $\Omega$ is finite.)
Fig. \ref{fig:exchange ratio} shows that for $ r_s \leq 80$,  at least four  processes, $J_3$,  $J_2,$ $J_4$ and $J_6$, are appreciable, yielding a highly frustrated generalized Heisenberg model.
To the best of our knowledge, the phase diagram of this $J_2$-$J_3$-$J_4$-$J_6$ model in the presence of the Aharonov-Bohm  and Berry phase  (\ref{eq:Aharonov-Bohm},\ref{eq:Berry phase},\ref{eq:AB and Berry}) has not been systematically explored.
By contrast, the simpler $J_2$-$J_3$-$J_4$ model for $B=\Omega=0$ (i.e., $\phi_a = \gamma_a = 0 $)  has been studied in \cite{kubo1997ground, momoi1997possible, motrunich2005variational, cookmeyer2021four, misguich1999spin}.
For a non-zero $B$ or $\Omega$, the $J_2$-$J_3$-$J_4$ model [$J_5=J_6 = J_6' =0$ in \eqref{eq:WC exchanges}] becomes \ksk{\footnote{\ksk{The effective spin Hamiltonian in the lattice model (as opposed to the continuum model of the present work) in the presence of a magnetic field is analogously derived in \cite{sen1995large, motrunich2006orbital}. The chiral spin interaction $\chi_{ijk}$ can also arise in metals as demonstrated in \cite{panigrahi2024spin}.}}}
\begin{fleqn}
    \begin{align}
    \label{eq:J2-J3-J4}
    &H_{\rm eff} 
    \!\approx\!   \Big  ( 4|J_2| \cos \Phi_2      \!-\! 4|J_3| \cos \Phi_3 \!+\! 5 |J_4| \cos \Phi_4   \Big  )
    \! \sum_{\langle i,j \rangle} \!{\bf S}_i \!\cdot\! {\bf S}_j 
    \nonumber 
    \\ 
     &
     -4  |J_3| \sin  \Phi_3 \!\!
    \sum_{\langle i,j,k \rangle \in 
     \Triangles
    } \!\!\! 
    \chi_{ijk}
    +2   |J_4| \sin  \Phi_4 \!\!
    \!\!
    \sum_{\langle i,j,k\rangle \in \TriWide } \!\!\! \chi_{ijk}   
    \nonumber \\
    &
    + |J_4| \cos \Phi_4 \! 
    \sum_{\llangle i,j\rrangle} \! {\bf S}_i \cdot {\bf S}_j 
    +4  |J_4| \cos  \Phi_4 \!\!\!\!
    \sum_{\langle i,j,k,l\rangle\in \Rhombus } \!\!\!\!
    h_{ijkl},\!
    \\
    &\Phi_a[B_0, \Omega(\v k)] \equiv \phi_a[B_0] + \gamma_a[\Omega(\v k)].
    \end{align}    
\end{fleqn}
Here, $\langle i,j \rangle$ and $\llangle i,j \rrangle $  denote   nearest- and next-nearest-neighbor bonds;
the chiral spin interactions $\chi_{ijk}$ act on equilateral triangles ($\Triangles$) and isosceles triangles with
$120^\circ$ apex angle ($\TriWide$), with  $i,j,k$ ordered in a counter-clockwise direction in each triangle];
the last term is the four-spin interaction \eqref{eq:h_ijkl} on each rhombus ($\Rhombus$), with $i,j,k,l$ ordered counter-clockwise in each rhombus.
The signs of these interactions  are tunable  via $\Phi_a$, which is a function of $B_0$, $\Omega(\v k)$ and the density.
Notably, chiral  interactions on $\Triangles$, next-nearest-neighbor Heisenberg exchanges and four-spin terms $h_{ijkl}$ are each known to stabilize the Kalmeyer-Laughlin chiral spin liquid (CSL)  \cite{wietek2017chiral, gong2017global, cookmeyer2021four}. 
Since all of them are present here, the CSL phase may arise over a broad  parameter range in a WC in the presence of $B$ and/or $\Omega$.

Recent experiments report an extended region of the WC phase in rhombohedral multilayer graphene (RMG) \cite{lu2025extended,lu2024fractional}, suggesting that our results may be directly relevant to the WC and proximate correlated phases in this system.
\ksk{For concreteness, we provide a rough estimate of the exchange energy scale using data from the tetra-layer device   \cite{han2025signatures}. We take the interlayer potential difference between the top and bottom layer to be $\Delta = 90 {\, \rm meV}$ (for which the band structure becomes most flat), the carrier density $n_e = 0.4 \times 10^{12} {\, \rm cm^{-2}}$ and the relative permittivity $\epsilon _r \approx 5$. From the band structure calculation shown in Fig. 12 (a) of \cite{han2025signatures}, the effective mass and Fermi energy at this density are roughly $m^*/m_e \approx 3$ and $E_F \approx 1 {\, \rm meV}$, respectively. These parameters yield an effective Hartree energy $E_h = \frac{m^*/m_e}{\epsilon_r^2 } (27.2 {\, \rm eV}) \approx 3.3 {\, \rm eV }$, an effective Bohr radius $a_B = \frac {\epsilon_r}{m^*/m_e} (5.29 \times 10^{-11}{\rm m}) \approx 8.8 \times 10^{-11} {\rm m}$ and an average inter-particle distance $r_0 = 1/{\sqrt {\pi n_e}}\approx 8.9 \times 10^{-9} {\rm m}.$ The resulting Coulomb energy scale is $E_{\rm Coulomb } \sim E_h \frac{a_B}{r_0} \approx 33{\, \rm meV}$, corresponding to an interaction parameter $r_s \equiv  E_{\rm Coulomb }/ E_F \sim 30\text{--}40.$ The effective exchange scale thus is $J_{\rm eff}\sim E_h \frac {e^{-\sqrt{r_s } S}}{r_s^{5/4}} \sim 0.1 \text{--}0.01 {\, \rm K},$ where we took $S \approx 1.5$ (Table \ref{table:results}). By contrast,  the valley ordering in the WC arises in the large-$r_s$ limit from the trigonal warping at order $V_{\rm eff} \sim E_h /r_s^{3}$ \cite{calvera2022WC}, which evaluates to $V_{\rm eff} \sim 1\, {\rm K}$. Therefore, in  rhombohedral graphene, valley ordering of the WC occurs at a temperature scale  orders of magnitude larger than that of spin ordering, $J_{\rm eff} \ll V_{\rm eff}$. Moreover, since the spin $g$-factor in RMG is close to the bare electron value $(= 2)$, a magnetic field larger than   $B \sim  \frac{J_{\rm eff}}{2 \mu_B} \sim 0.1 \text{--} 0.01 {\, \rm T}$ is sufficient to  fully polarize the spin degrees of freedom.

Taken together, our work shows how a magnetic field and Berry curvature modify the magnetic properties of a quantum crystal, in particular by generating chiral spin terms that can stabilize the chiral spin-liquid phase. More broadly, our work highlights that the effects of those geometric phases manifest fundamentally differently in strongly correlated continuum systems as compared to weakly interacting systems or Hubbard-type lattice models.
}

{\it Note added:} Upon completion of this work, Ref. \cite{joy2025chiral} appeared, which partially overlaps with this work. 
The results are largely in agreement where they overlap.
\\

\section*{Acknowledgment}
I am grateful to Pavel Nosov for insightful discussions that motivated this investigation, and to Steve Kivelson, Liang Fu, Michael Stone, Long Ju, Daniel Arovas, and especially Eduardo Fradkin for valuable discussions. 
I also gratefully acknowledge Eduardo Fradkin and Ilya Esterlis for useful comments on the draft.
This work  benefited from discussions with  participants of the Joint ICTP-WE Heraeus School and Workshop on Advances on Quantum Matter, my travel to which was partially supported by the US National Science Foundation (NSF) Grant No. 2201516 under the Accelnet program of Office of International Science and Engineering
(OISE).
I am supported by the Anthony J. Leggett Postdoctoral Fellowship at the University of Illinois  Urbana-Champaign.
This work was  performed in part at the Aspen Center for Physics, which is supported by a grant from the Simons Foundation (1161654, Troyer).
Some of the computing for this project was performed on the Sherlock cluster at  the Stanford Research Computing Center.

\bibliography{ref.bib}

\begin{thebibliography}{76}
\bibitem [{foo({\natexlab{c}})}]{footnote:modulation}%
  \BibitemOpen
  \bibinfo {note} {The magnitude of the modulating piece $\delta B(\v r)$ and $\delta \Omega(\v k)$ are assumed to be of the same order as $B_0$ and $\Omega_0$, respectively.}\BibitemShut {Stop}%
\bibitem [{foo({\natexlab{d}})}]{footnote:LL-filling-factor}%
  \BibitemOpen
  \bibinfo {note} {We note that $\nu$ here should not be interpreted literally as a filling factor of a Landau level, since we work in the limit where the Coulomb interaction is the dominant energy scale and hence the Landau-level mixing is substantial.}\BibitemShut {Stop}%
\end{thebibliography}

\end{document}